\documentclass[a4paper,fleqn,usenatbib,useAMS]{mnras}
\usepackage{graphics}
\usepackage{graphicx}
\usepackage[pagewise]{lineno}
\usepackage{amsmath}    
\usepackage{amssymb}    
\usepackage{multicol}        
\usepackage{bm}     
\usepackage{pdflscape}  
\usepackage{xcolor}
\usepackage{ae,aecompl}
\usepackage{subfigure}
\usepackage{scalerel}
\usepackage[english]{babel}
\usepackage{times}
\usepackage{physics}
\usepackage{etoolbox}
\usepackage{physics}
\usepackage{etoolbox}
\usepackage{multirow, booktabs}
\usepackage{booktabs,siunitx}
\usepackage{siunitx}
\usepackage{paralist}
\usepackage{makecell}
\usepackage{tabularx}
\usepackage{paralist}
\usepackage[inline]{enumitem}
\newlist{mycompactenum}{enumerate}{1}
\setlist[mycompactenum,1]{nosep,label=\arabic*.}

\newcommand{\rrl}{RR~Lyrae~}
\newcommand{\rrls}{RR~Lyraes~}

\title[Microlensing of low-amplitude pulsating stars]{Identifying low-amplitude pulsating stars through microlensing observations}
\author[Sajadian et al.]{Sedighe Sajadian$~^{1,2}$\thanks{E-mail: s.sajadian@iut.ac.ir}, Richard Ignace$~^{3}$\thanks{E-mail: ignace@etsu.edu}, Hilding Neilson$~^{4}$\thanks{E-mail: neilson@astro.utoronto.ca}\\
$^{1}$Department~of~Physics,~Isfahan~University~of~Technology,~Isfahan~84156-83111,~Iran\\
$^{2}$Department~of~Physics,~Chungbuk~National~University,~Cheongju~28644,~Republic~of~Korea\\
$^{3}$Department of Physics \& Astronomy, East Tennessee State University, Johnson City, TN 37614, USA \\
$^{4}$David A.~ Dunlap Department of Astronomy \& Astrophysics, 50~St.~George Street, University of Toronto, Toronto, ON, M5S 3H4, Canada}

\date{Accepted XXX. Received YYY; in original form ZZZ}
\pubyear{2021}

\begin{document}
	
	
\label{firstpage}
\pagerange{\pageref{firstpage}--\pageref{lastpage}}
\maketitle
\begin{abstract}
One possibility for detecting low-amplitude pulsational variations is
through gravitational microlensing.  During a microlensing event,
the temporary brightness increase leads to improvement in the
signal-to-noise ratio, and thereby better detectability of pulsational
signatures in light curves.  We explore this possibility under two
primary considerations.  The first is when the standard point-source
and point-lens approximation applies.  In this scenario, dividing the
observed light curve by the best-fitted microlensing model leads to residuals that result in pulsational 
features with improved uncertainties.
The second is for transit events (single lens) or caustic crossing
(binary lens).   The point-source approximation breaks down, and
residuals relative to a simple best-fitted microlensing model display
more complex behavior. We employ a Monte-Carlo simulation of
microlensing of pulsating variables toward the Galactic bulge for the surveys of OGLE and of KMTNet. We demonstrate
that the efficiency for detecting pulsational signatures with
intrinsic amplitudes of $<0.25$ mag during single and binary
microlensing events, at differences in $\chi^{2}$ of $\Delta \chi^{2} >350$, is $\sim 50-60\%$. The maximum
efficiency occurs for pulsational periods $P \simeq 0.1-0.3$ days.
We also study the possibility that high-magnification microlensing
events of non-radially pulsating stars (NRPs) could be misinterpreted as
planetary or binary microlensing events. We conclude that small
asymmetric features around lightcurve peaks due to stellar pulsations
could be misdiagnosed with crossing (or passing close to) small
caustic curves.

\end{abstract}

\begin{keywords}
gravitational lensing: micro, stars: oscillations (including pulsations), methods: numerical, software: simulations
\end{keywords}

\section{Introduction}\label{two}

Variable stars are classified into two main categories: extrinsic and intrinsic ones. The extrinsic variables are either eclipsing binaries or rotating variables. The variability features for stars in the second class are due to their interior evolution which changes their luminosity with time. Two main subclasses of intrinsic periodic variables include \rrl and Cepheids. The \rrl stars have periods in the range of a few hours to a few days. These variable stars are mostly sub-giants, located on the horizontal branch, and frequently found in globular clusters. They are similar to the classical Cepheid variables, but with shorter periods and lower luminosities \citep[see, e.g., ][]{Cox1974, Kahn1969, book2007}. Cepheid variables have periods in the range of $P \in [2,~60]$ days and luminosities $L \sim [300,~40000]L_{\odot}$, whereas \rrls have a lower average luminosity of around $80~L_{\odot}$. Indeed, they are population~II stars with low-metallicity values. While frequently observed in globular clusters, they are also common in other old stellar populations, such as the Galactic bulge. \rrl stars are often used as distance indicators as well as tracers of old stellar populations \citep[see, e.g.,][]{Jameson1986,Smith1984}. Because of the faintness of \rrls, they are challenging to detect as extragalactic sources, or within the Galaxy when of small-amplitude pulsations.

In any given stellar survey some variables go undetected, either because of the intrinsic faintness of the stars (i.e., low luminosity or great distance), or owing to the low amplitude of the variations relative to the detection threshold for the survey. Among currently operating ground-based surveys that have produced large numbers of variable star detections is the Optical Gravitational Lensing Experiment (OGLE, \citet{OGLE_IV}). The OGLE group has discovered more than $56,000$ \rrl stars toward the Galactic disk and bulge, and many others towards the Magellanic clouds, using the $1.3$m Warsaw telescope. Most have amplitudes larger than $0.1$ mag \footnote{\url{http://ogledb.astrouw.edu.pl/~ogle/OCVS/catalog_query.php}} \citep[e.g., ][]{OGLERR1, OGLERR3, OGLE2018Chefeid}. In 2019, the Gaia second data release published a catalog of variable stars throughout the whole sky  with amplitudes larger than $0.1$ mag \citep{Gaia2019RR}. Lower-amplitude \rrl variables are barely detectable through these photometric observations, but could be more easily and confidently identified if the stars’ brightnesses were magnified through gravitational microlensing. Microlensing produces a temporary enhancement in the stellar brightness when closely aligned with a foreground and object of sufficient mass  \citep{Einstein1936, Liebes1964, Chang1979, Pac1986, sajadian665}.

In 1997, the first microlensing event of a variable source star was discovered towards the Small Magellanic Cloud (SMC), namely $\rm{MACHO}$-$97$-$\rm{SMC}$-$1$ \citep{alcock1997}. In this event, the blending-parallax degeneracy was resolved using the variability feature of the source star \citep{Assef2006}. Another microlensing candidate from a variable and bright source star was reported recently. For this event, the asteroseismic analysis was done to find the intrinsic variability curve of the source star from the baseline data \citep{Varmicro}. The OGLE collaboration announced 137 candidates of microlensing events of variable source stars \citep{Lukaz2006}. According to their results, for $10\%$ of total detected microlensing events, the source stars show variability signatures at their baselines.  Here, ``baseline'' refers to the lightcurve as unaffected by the lensing, i.e., before and after the event. A non-variable star would have a flat baseline light curve.  The OGLE result indicates that a significant fraction of total microlensing events involve variable stars and hints at there being more low-amplitude variables that went undetected.

The theoretical framework of microlensing involving NRPs was first introduced in \citet{Bonanno2004}. Recently, the characterizations of (single and binary) microlensing lightcurves from radial and non-radial pulsating stars have been studied extensively in \citet{PaperI, PaperII}. As an extension of these two papers, this study explores the following two issues pertaining to microlensing of variable stars.  (1) How gravitational microlensing can enhance the detectability of these low-amplitude pulsating stars.  (2) How microlensing lightcurves of these low-amplitude variables could in some cases masquerade and be misinterpreted as other second-order effects associated with microlensing of non-variable stars.  We conduct Monte-Carlo simulations of single and binary microlensing events of radially and non-radially pulsating variable stars and generate synthetic light curves relevant to the OGLE \citep{OGLEIVphase} and the Korea Microlensing Telescope Network (KMTNet, \citet{KMTNet2016}) surveys.  A description of our simulations are given in section \ref{simul}. Then, in section \ref{varib}, we study the efficiency for detecting stellar variability during microlensing events. In section \ref{miss}, we explore the conditions under which microlensing lightcurves of low-amplitude and non-radially pulsating variables
may be degenerate with planetary/binary lightcurves. Finally, we summarize our conclusions in section \ref{conclu}.

\section{Simulating Microlensing of low-amplitude variables}\label{simul}

We explain the details of our Monte-Carlo simulation of  single and binary microlensing events toward the Galactic bulge in subsection \ref{firstsub}. We assume that source stars in the simulation have some low-amplitude pulsation features. In order to study their detectability, we generate synthetic data points for simulated events based on the OGLE and KMTNet surveys as explained in subsection \ref{thirdsub}.

\subsection{Simulating microlensing events}\label{firstsub}

We have constructed a source star population using the Galactic Besan\c{c}on model \footnote{\url{https://model.obs-besancon.fr/}} for luminosity classes I,~II,~III,~IV and V (supergiants through main-sequence stars) \citep{Robin2003,Robin2012} with the same method as explained in \citep{sajadian2019, Moniez2017, 2015sajadian}. The distances of the source stars from the observer ($D_{\rm s}$) are determined using the cumulative mass versus distance for each given direction $dM/dD_{\rm s} (l,~b)$, where $(l,~b)$ represent the Galactic longitude and latitude, respectively \citep[see, e.g.,][]{sourcedis}.

We have generated another stellar population for the lens stars. The distance of each lens with respect to the observer ($D_{\rm l}$) is taken from the microlensing event rate versus distance $d\Gamma/dD_{\rm l} \propto \rho_{\rm t}(D_{\rm l},~l,~b)~R_{\rm E}$. Here, $\rho_{t}$ is the cumulative mass density in the given direction and distance from the observer, and $R_{\rm E}$ is the Einstein radius (the ring of the image for perfect alignment).

Our simulation is pertinent for the Galactic bulge over a region with Galactic longitude $l\in [3,~6^{\circ}]$ and Galactic latitude $b \in [-6,~-4^{\circ}]$. We exclude events within which the baseline magnitude of the source stars (by considering the blending effect) is fainter than $22$ mag in the $I$-band. Events with Einstein crossing times longer than $300$ days are also ignored.

For binary microlensing events, the mass ratio of the two stars and the angular separation of the components are uniformly sampled from the ranges $q \in [0.4,~1.0]$ and $d \in [0.6,~1.4]$, where $d$ is the projected angular separation of the lenses on the sky as normalized to the angular Einstein radius. We evaluate the magnification factor in binary microlensing events with the well-developed $\rm{RT}$-model \footnote{\url{http://www.fisica.unisa.it/gravitationastrophysics/RTModel.htm}} by V.\ Bozza \citep{Bozza2018,Bozza2010,Skowron2012}. The lens angular impact parameter in the sky, normalized to the angular Einstein radius, is also sampled uniformly, in the range $u_{0}\in [0,~1]$.
\begin{figure*}
	\centering
	\includegraphics[width=0.49\textwidth]{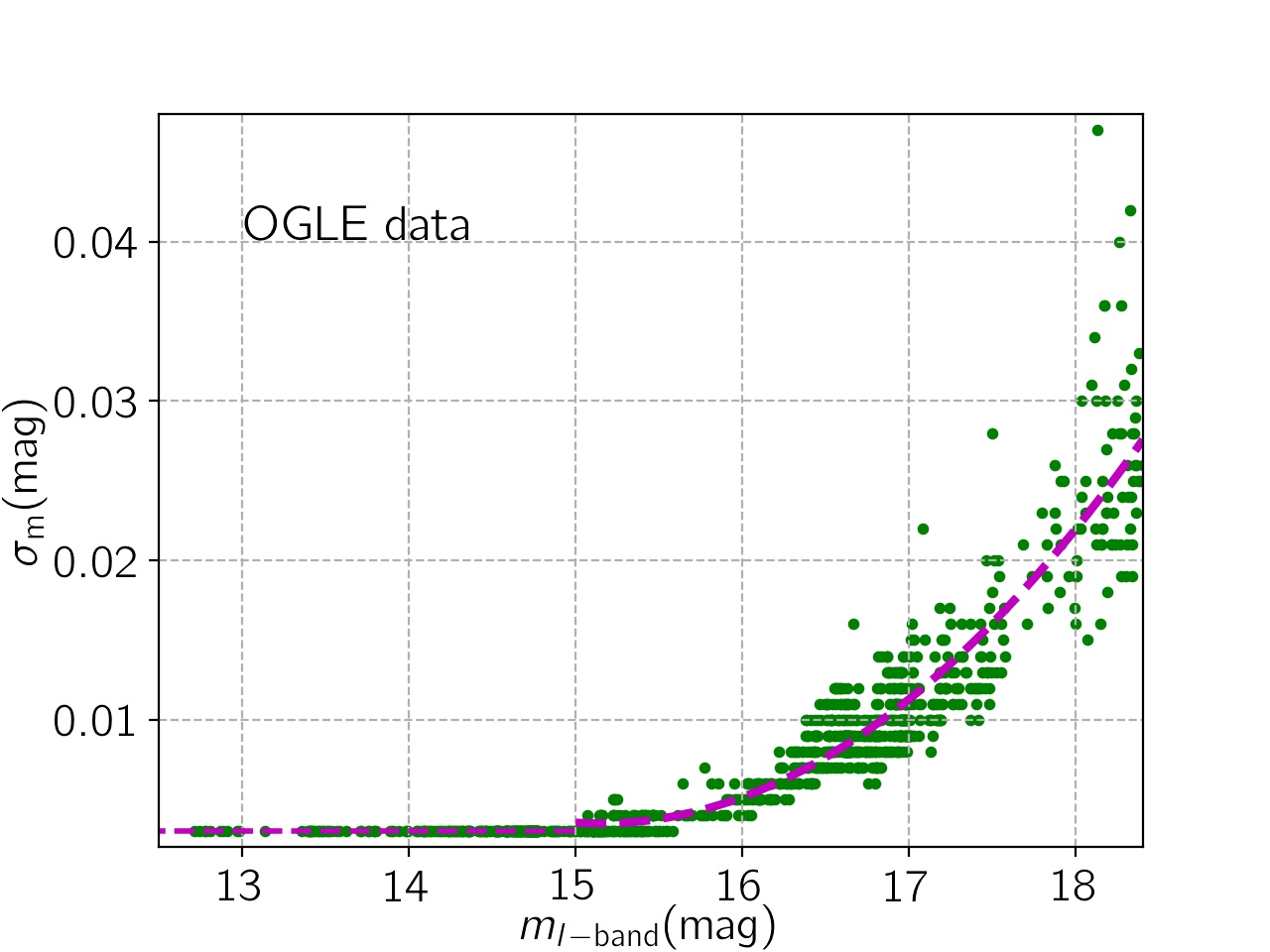}
	\includegraphics[width=0.49\textwidth]{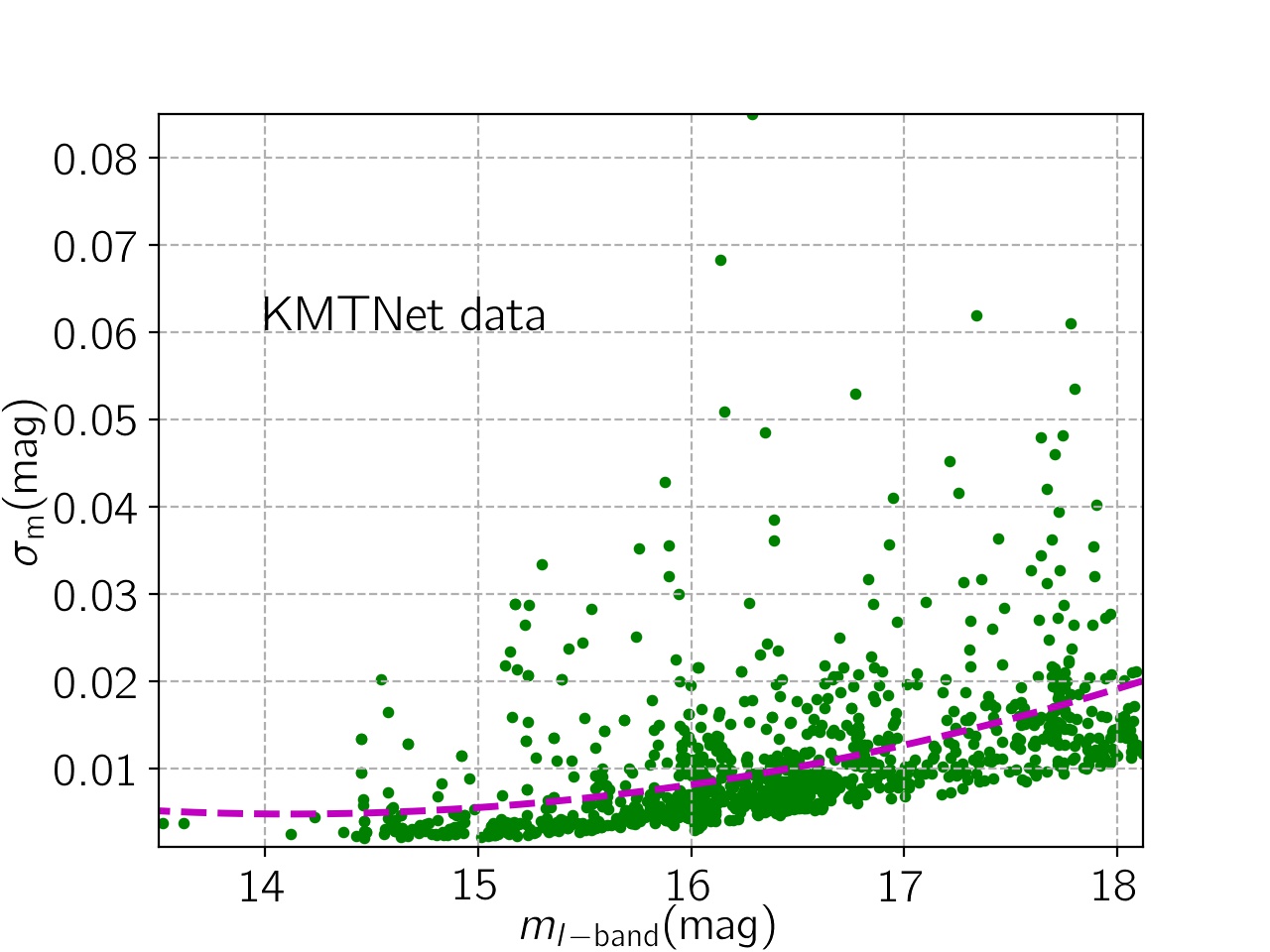}
	\caption{The photometric uncertainties (in magnitudes) of data points for single and binary microlensing events as seen by OGLE (left panel) and KMTNet (right panel) versus $I$-band apparent magnitude of the source stars. The dashed magenta lines are the best-fitted polynomial curves to the observations.} 
	\label{errorpd}
\end{figure*}

\subsection{Pulsational features}\label{secondsub}
To differing degrees, all stars are variable. Variations in the brightnesses of intrinsic pulsating stars are modeled with either simple sinusoidal functions to represent radially pulsating stars, or spherical harmonic functions to represent non-radially pulsating stars. Details  for simulating microlensing of such stars appears in \citet{PaperI, PaperII}, an approach that we also adopt here. The simulation of radially pulsating stars requires two parameters: the pulsation period $P$ and brightness amplitude $\delta_{\rm m}(\rm{mag})$. For these stars, the radius and surface temperatures vary periodically. In the case of non-radially pulsating stars, further parameters are needed: the inclination angle $i$ between the plane of the sky and the polar axis of the star, and two numbers for determining the harmonic modes of the stellar pulsation $l,~m$.

Then for our simulations, stellar pulsation periods are randomly selected uniformly in the range $P \in [1~\rm{hr},~15~\rm{days}]$.  In order to simulate low-amplitude pulsating stars, we randomly select the magnitude amplitude in the $I$-band as uniform in the range $\delta_{\rm m} \in [0.005,~0.25]$ mag. We note that most detected variable stars have amplitudes larger than $0.1$ mag. Hence, our simulations include variables of lower amplitude than is detectable at the baseline of the OGLE and KMTNet surveys.

\subsection{Synthetic data points}\label{thirdsub}
To evaluate the detectability of pulsation properties from microlensing light curves, we have generated synthetic lightcurves relevant to groups that currently conduct large surveys of stellar populations. We have produced ensembles of cadences and photometric uncertainties to replicate characteristics of single and binary microlensing events of giant stars that have been detected with OGLE \footnote{\url{http://ogle.astrouw.edu.pl/ogle4/ews/ews.html}}
and KMTNet\footnote{\url{http://kmtnet.kasi.re.kr/ulens/}} programs \footnote{We do not include the The Microlensing Observations in Astrophysics (MOA, \citet{MOA_gourp}) observations, because of its filter $R$-band used to observe microlensing events. The microlensing light curves from variable stars are chromatic. These lightcurves in different passbands do not have similar shapes, so we can not convert these data with just shifting.} The photometric uncertainties depend on the source brightness, with better uncertainties for brighter sources. In the simulations photometric uncertainties are
determined from the source apparent magnitude. Figure \ref{errorpd} shows a scatter plot of photometric uncertainties in magnitude units versus the source magnitude for data from OGLE (left panel) and KMTNet (right panel) surveys. We fit two second-order polynomial functions to the data sets, as shown by the magenta dashed lines. For the OGLE survey, the photometric uncertainty is $\sim 0.003$ mag for stars brighter than $15$ mag in $I$-band; for fainter stars, the average uncertainty is given by:

\begin{equation}
\sigma_{m} =0.01\left [0.86+0.68(m_{I}-16.7)+0.23(m_{I}-16.7)^{2}\right ].
\end{equation}

\noindent For KMTNet data, the uncertainty for stars brigher than $14$ mag in the $I$-band is $\sim 0.0038$ mag; for fainter sources, the average uncertainty is given by: 

\begin{eqnarray}
\sigma_{m} =0.01\left [7.4-1.6(m_{I}-5.6)+0.1(m_{I}-5.6)^{2}\right ].
\end{eqnarray}

\noindent The above trends for photometric errors are incorporated into our simulations for $I$-band light curves with microlensing. Although the microlensing monitoring surveys are conducted in the $I$-band filter, we additionally calculate magnification light curves in the $BVR$-bands.

The time interval between data points also mimics actual observations. We downloaded observational data for several lightcurves taken by the OGLE and KMTNet surveys. Accordingly, we made two ensembles of the observing cadences. While simulating sythetic data points, their cadence is determined from those ensembles.

\begin{figure*}
\centering
\subfigure[]{\includegraphics[width=0.48\textwidth]{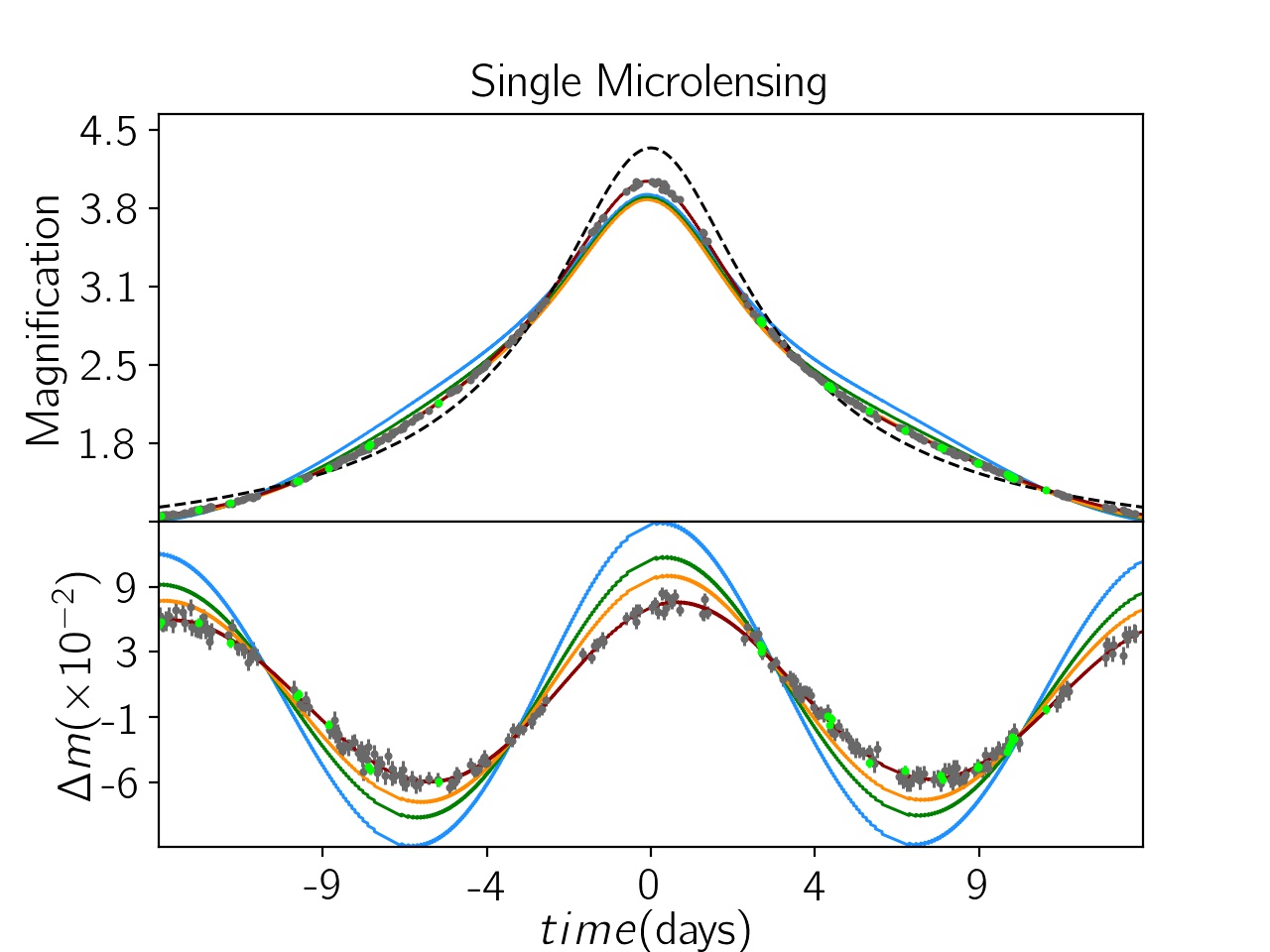}\label{fig2a}}
\subfigure[]{\includegraphics[width=0.48\textwidth]{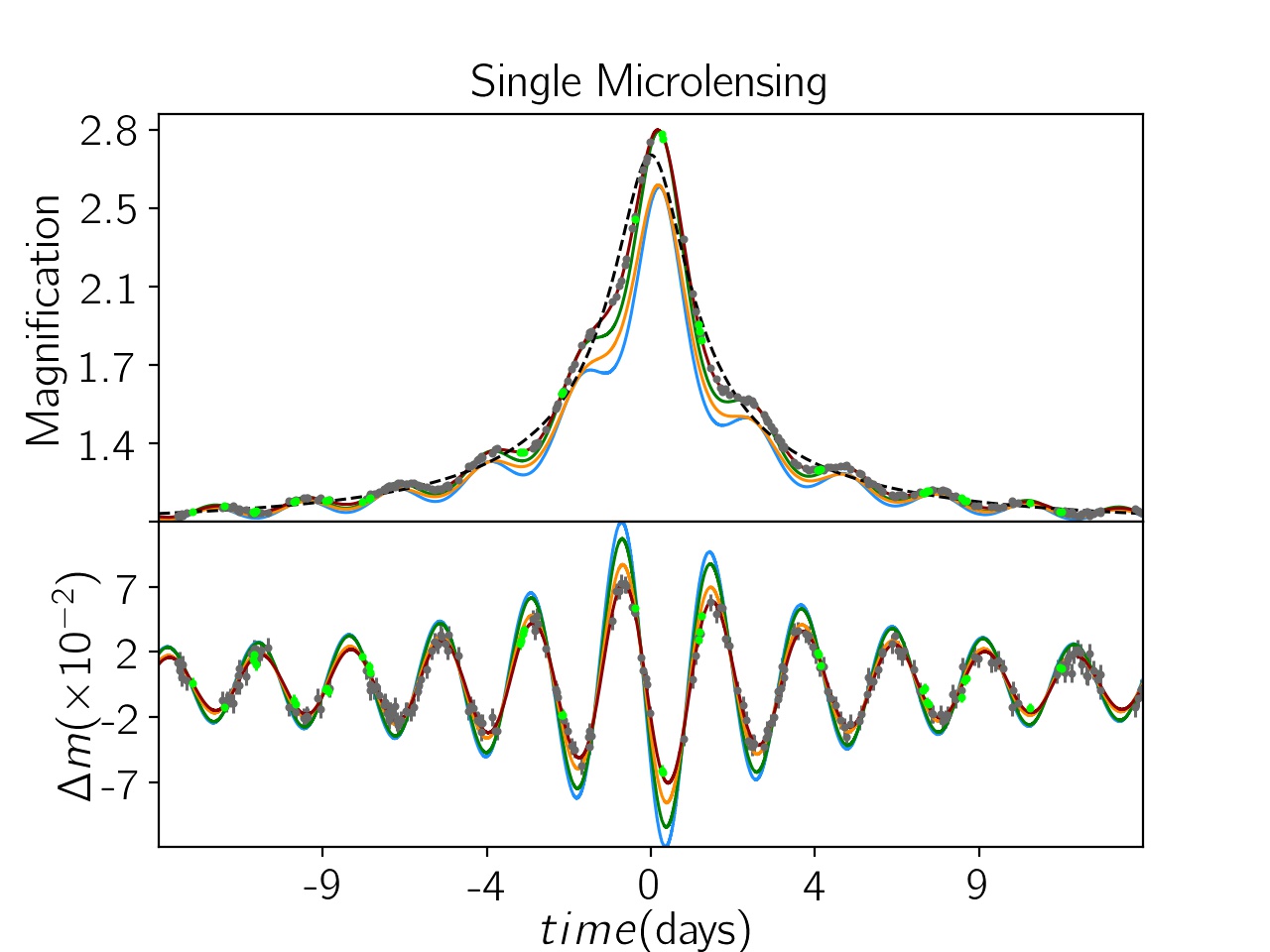}\label{fig2b}}
\subfigure[]{\includegraphics[width=0.48\textwidth]{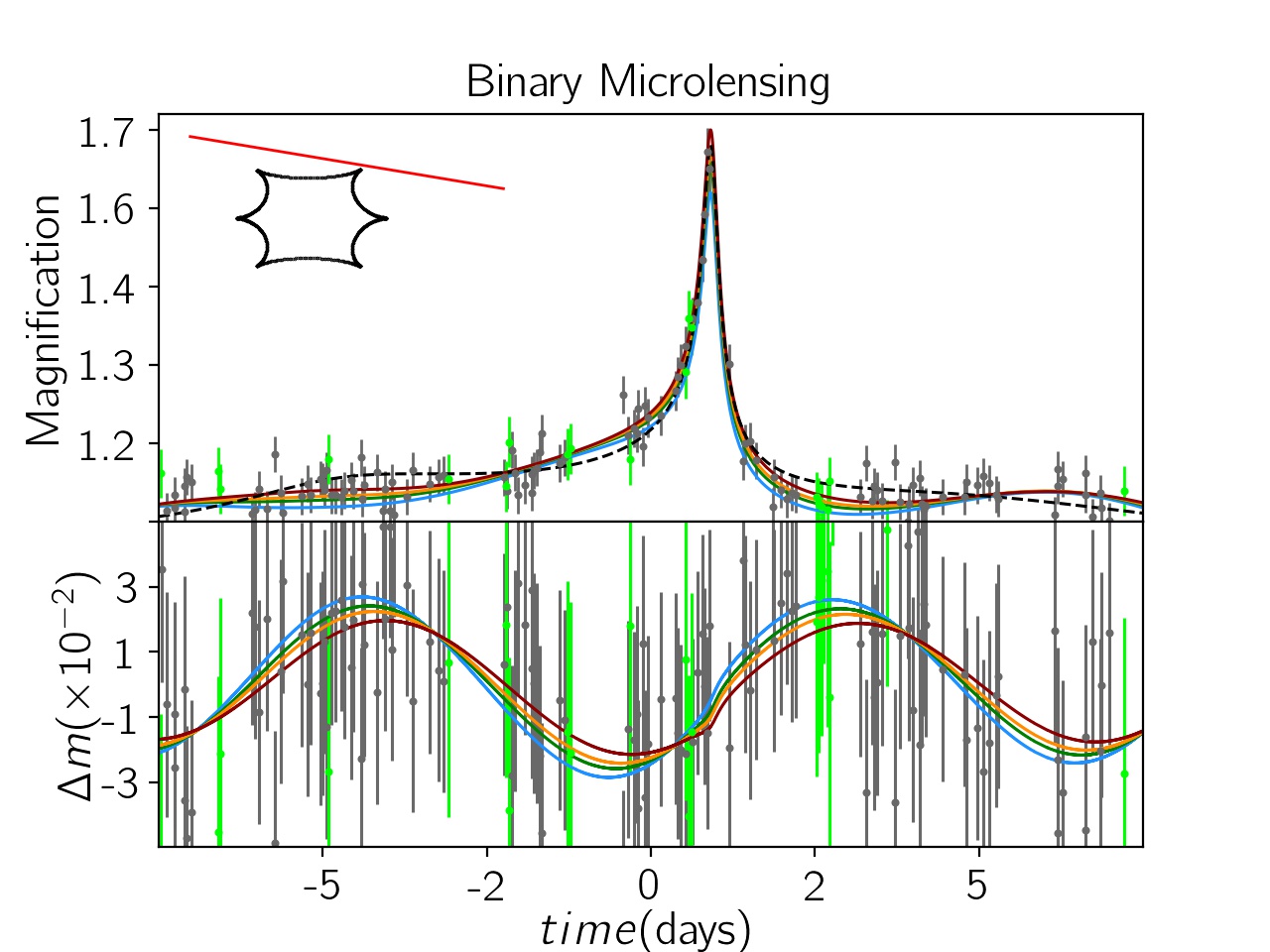}\label{fig2c}}
\subfigure[]{\includegraphics[width=0.48\textwidth]{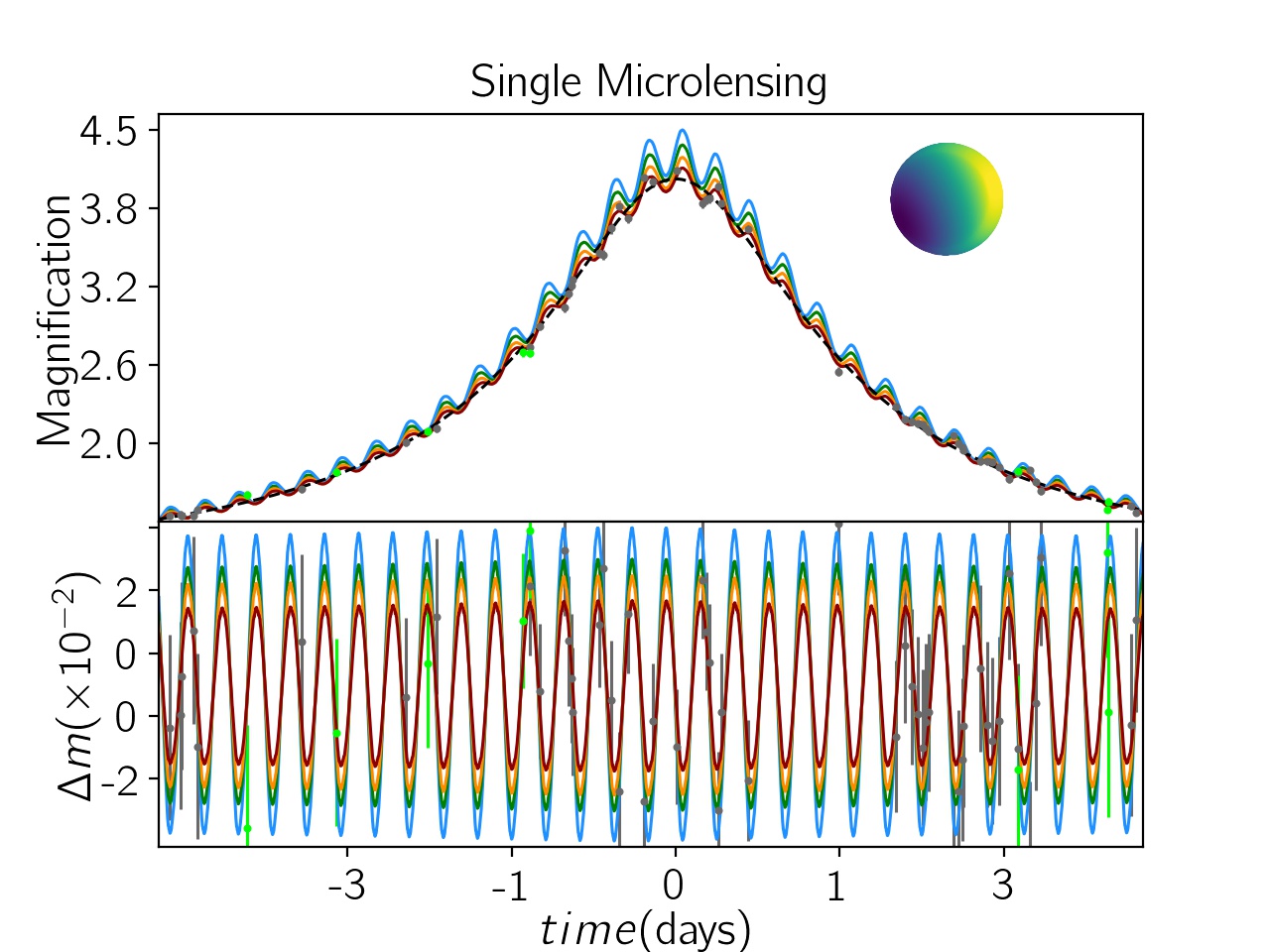}\label{fig2d}}
\subfigure[]{\includegraphics[width=0.48\textwidth]{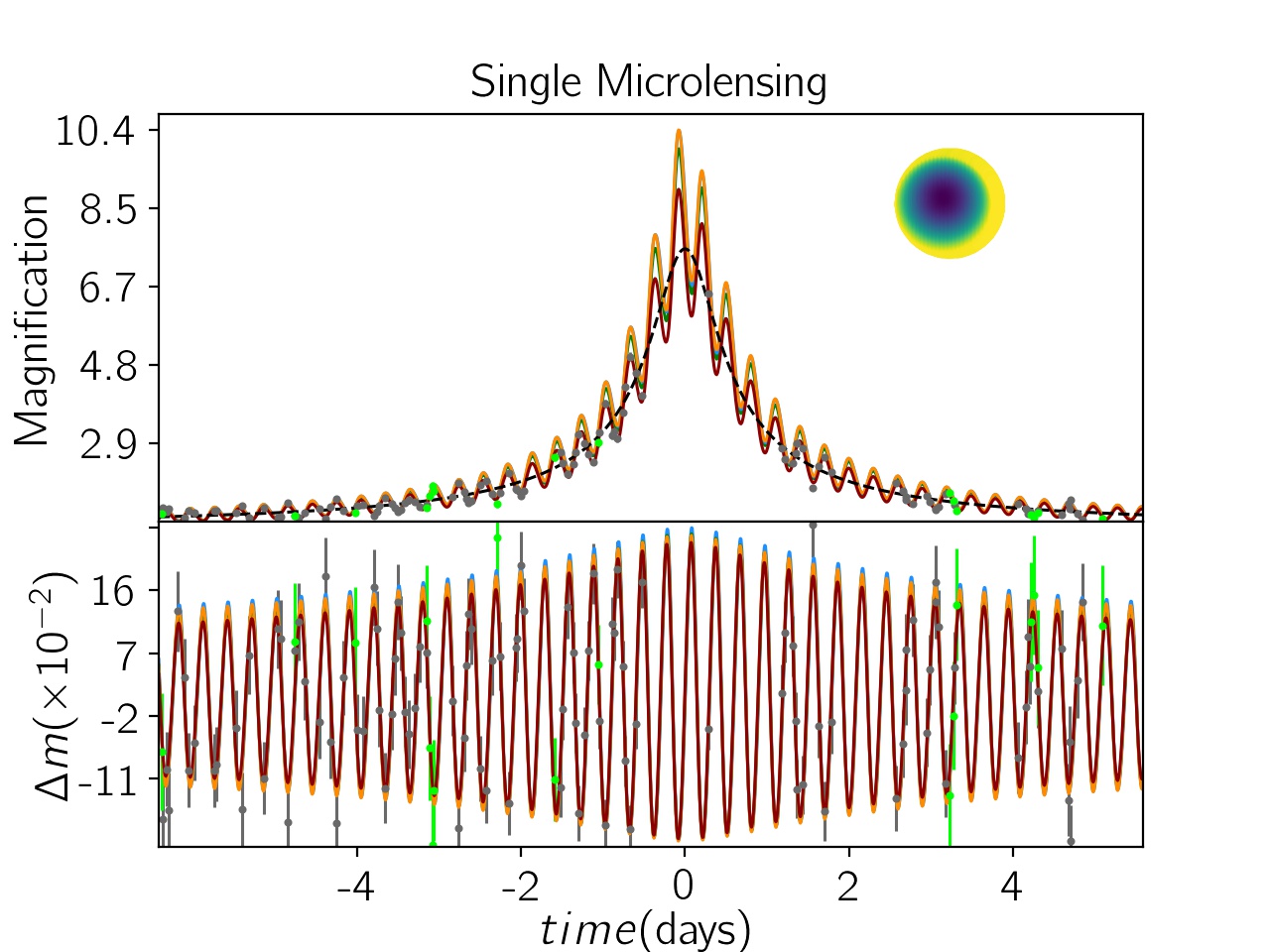}\label{fig2e}}
\subfigure[]{\includegraphics[width=0.48\textwidth]{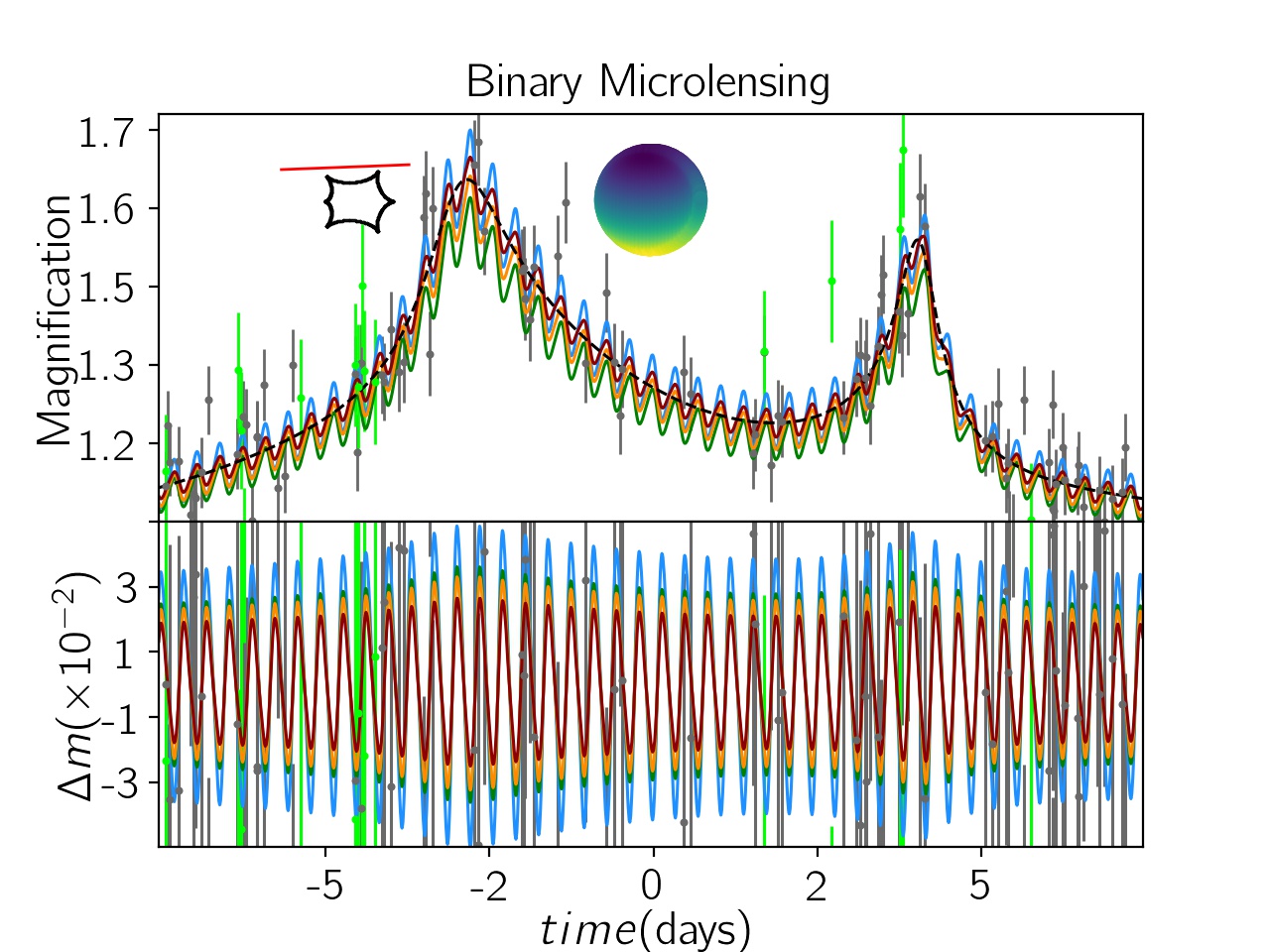}\label{fig2f}}
\caption{Examples of single and binary microlensing light curves of pulsating stars. Lightcurves (magnification factor) are for standard $BVRI$-bands as blue, green, orange and dark-red colors, respectively. The black dashed  curves represent the calculated microlensing light curves without pulsation. Synthetic data points are plotted with uncertainties and cadences appropriate for OGLE (lime) and KMTNet (grey) surveys. The residuals display differential magnitudes in $BVRI$-bands due to the stellar pulsation, given by Equation \ref{resi}. The parameters of the light curves can be found in Tables~\ref{tab1} and \ref{tab2}. For binary events (Fig. \ref{fig2c} and \ref{fig2f}), the caustic curves (black curves) and the source trajectories over the lens plane (red lines) are shown in insets. In the last three events, the source stars are non-radially pulsating variables, and their surface temperature variations are represented in insets with colored schemes.}
\label{lightcurve}
\end{figure*}

\begin{table*}
	\centering
	\caption{Parameters of microlensing lightcurves shown in Figure \ref{lightcurve} from radially pulsating stars. Here, $m_{\rm{base}}$ is the apparent magnitude due to the average stellar flux over the baseline and $\left<\sigma_{\rm{m}}\right>/\left<\sigma_{\rm{base}}\right>$ is the ratio of the averaged photometric error bar during lensing to that from the baseline amount. The last column indicates if the stellar pulsations are detectable during microlensing (labeled by $\rm{D}$) or not (labeled by $\rm{ND}$).}
	\begin{tabular}{cccccccccccc}\toprule[1.2pt]
		$\rm{Fig.~No.}$& $\delta_{\rm m}$ & $P$ & $u_{0}$ & $\rho_{\star}$ & $t_{\rm{E}}$ & $m_{\rm{base}}$ & $\Delta \chi^{2}_{\rm{lens}}$& $\left<\sigma_{\rm{m}}\right>/\left<\sigma_{\rm{base}}\right>$  & $q$ & $d$ &  $\rm{ND}/\rm{D}$\\
		& $\rm{(mmag)}$  & $\rm{(days)}$  & & &$\rm{(days)}$ & $\rm{(mag)}$ & & & & $\rm{(R_{E})}$ & \\	
		\toprule[1.2pt]
		\ref{fig2a} & $83.58$ & $14.34$ & $0.17$ & $0.001$ & $15.4$ & $14.8$ & $38170$ & $0.9$ & $--$ & $--$ & $\rm{D}$\\
		\ref{fig2b} & $124.13$ & $2.52$ & $0.07$ & $0.001$ & $15.2$ & $14.9$ & $11182$ & $0.9$ & $--$ & $--$ & $\rm{D}$\\
		\ref{fig2c} & $75.47$ & $7.79$ & $0.65$ & $0.002$ & $8.9$ & $19.3$ & $42$ & $1.0$ & $0.9$ & $1.1$ & $\rm{ND}$\\
		\hline
	\end{tabular}
	\label{tab1}
\end{table*}
\begin{table*}
	\centering
	\caption{The parameters of microlensing lightcurves shown in Figure \ref{lightcurve} from non-radially pulsating stars.}
	\begin{tabular}{ccccccccccccccc} \toprule[1.2pt]
		$\rm{Fig.~No.}$& $\delta_{\rm m}$ & $P$ &  $i$  & $l$ & $m$ & $u_{0}$ & $\rho_{\star}$ & $t_{\rm{E}}$ & $m_{\rm{base}}$ & $\Delta \chi^{2}_{\rm{lens}}$& $\left<\sigma_{\rm m}\right>/\left<\sigma_{\rm{base}}\right>$ & $q$ & $d$ & $\rm{ND}/\rm{D}$ \\
		& $\rm{(mmag)}$  & $\rm{(days)}$ & $\rm{(deg)}$ & & & & &$\rm{(days)}$ & $\rm{(mag)}$ & & & & $\rm{(R_{E})}$ &\\	
		\toprule[1.2pt]
		\ref{fig2d} & $25.62$ & $9.6$ & $47.8$ & $1$ & $1$ & $0.22$ & $0.002$ & $5.8$ & $18.4$ & $95.5$ & $0.7$ & $--$ & $--$ &  $\rm{ND}$\\
		\ref{fig2e}& $265.97$ & $4.05$ & $82.5$ & $2$ & $0$ & $0.06$ & $0.001$ & $8.3$ & $20.5$ & $1151$ & $0.8$ & $--$ & $--$ &  $\rm{D}$\\ 
		\ref{fig2f}& $44.04$ & $8.45$ & $22.6$ & $1$ & $0$ & $0.49$ & $0.001$ & $8.4$ & $21.6$ & $5.9$ & $0.9$ & $0.4$ & $1.3$ &  $\rm{ND}$\\
		\hline
	\end{tabular}
	\label{tab2}
\end{table*}

For simulating light curves, we have primarily three key timescales for a single lens, and four when considering a binary lens.  The foreground lens and background source have a relative proper motion, $\mu_{\rm rel}$, with respect to each other.  The lens has a characteristic angular Einstein radius, $\theta_{\rm E}$ (which is the angular radius of the images ring for the prefect alignment),  and the source has an angular stellar radius, $\theta_\ast$. One key timescale is the crossing of the Einstein radius, as given by $t_{\rm E} = \theta_{\rm E}/ \mu_{\rm rel}$.  This sets the duration of the microlensing event  \citep[e.g., ][]{gaudi2012}. The next timescale is the crossing time of the star, $t_\ast =\theta_\ast/\mu_{\rm rel}$.  For source transits or near transits by the lens, $t_\ast$ sets the timescale for which finite source effects are relevant and for the polarimetry signals in microlensing events \citep{1994wittmoa, 1995MNRAS.276..182S, Sajadian2016, Sajadian2019b}.  Then for a  variable star, the pulsation period is the third relevant timescale. For a binary lens, the fourth timescale is the orbital period \citep[e.g., ][]{2014sajadian}.  For reference, typical values of $t_{\rm E}$ can be months, $t_\ast$ can be hours to days, and $P$ can be hours to days for pulsating stars. With these fiducials in mind, the orbital period for a binary lens is not likely relevant unless of order a year or less, since the orientation of the mass  components will be fixed in the sky for longer periods. With these scales in mind, we have computed synthetic lightcurves in the time interval $[-0.9,~0.9] ~t_{\rm E}$. For convenience, the time of closest approach is set to zero.

Figure \ref{lightcurve} shows examples of six simulated light curves. The first three, (a)--(c), are for radial pulsators, whereas the last three, (d)--(f), the source stars are for non-radial pulsators. Tables~\ref{tab1} and \ref{tab2} list model parameters specific to these plots.  In the figures, the black dashed curves represent the calculated microlensing light curve without pulsation. Light curves in the $BVRI$ bands with pulsations are shown in blue, green, orange and dark-red, respectively.  Measurement uncertainties are represented as data points with black for the KMTNet survey and lime for the  OGLE survey. In each of the six examples shown (a)--(f), the upper panel is the light curve, and the lower panel shows residuals when the black-dashed curve (i.e., for a non-pulsating star) is subtracted. 

Plots for (c) and (f) are for binary lens cases, and these display inset geometrical figures, with the caustic curve and the source trajectory shown as black curves and red straight lines, respectively.  For the case of NRPs, the pattern of surface temperature variations (i.e., $l,~m$ modes) are also shown as insets.

All the simulated light curves are shown in terms of the magnification factor, $A$. The magnification of the source stars is the modeled flux, $F_{\star}(t)$, normalized to the flux average at the baseline. This means that for a non-varying star with no microlensing, the light curve would ideally be a horizontal line with magnification of unity. The differing timescales along the horizontal axis signify the differing values of $t_{\rm E}$.  

\noindent The residuals are residual fluxes which are displayed in differential magnitude. To obtain the residuals, we find a best-fit magnification model for a non-pulsating star, $A_{\rm{best}}(t)$, and then calculate the differential magnitude for the residuals with

\begin{eqnarray}
\Delta m(mag)= -2.5 \log_{10}\left[\frac{F_{\star}(t)}{\left<F_{\star,\rm{base}}\right>~A_{\rm{best}}(t)}\right],
\end{eqnarray}\label{resi}

\noindent where, $\left<F_{\star, \rm{base}}\right>$ is the average stellar flux over the baseline entering the source PSF. The real microlensing lightcurve of a non-pulsating source star ($A_{\rm{best}}(t)$) is shown by the black dashed curves in the magnification panels. Again, all the parameters adopted for these lightcurves are given in Tables~\ref{tab1} and \ref{tab2}. In these tables, $m_{\rm{base}}$ is the apparent magnitude due to the average stellar flux over the baseline, and $\left<\sigma_{\rm{m}}\right>/\left<\sigma_{\rm{base}}\right>$ is the ratio of the averaged photometric error bar during lensing to that from the baseline amount (see the next subsection). For binary events (Fig. \ref{fig2c} and \ref{fig2f}), the caustic curves (black curves) and the source trajectories over the lens plane (red lines) are shown in insets. In the last three events, the source stars are non-radial pulsating variables, and their surface temperature variations  are represented with insets using the colored schemes.

Unless the lens transits or nearly transits the source star, the magnification curves from microlensing of pulsating stars derive from a multiplication of a simple magnification curve for microlensing of a non-varying star and the variability curve for the pulsations. For example, a classic result from the point-lens, point-source approximation for microlensing is the absence of chromatic signatures. This means that the residuals from the various filter bands will all be in phase.  In cases of transits or near transits, the residuals will not generally be strictly periodic, as for example, see the residual plots of Figure \ref{fig2b}.
Such effects occur when the point-source approximation is no longer valid, and persist in the light curve over the relatively short time scale of order $t_{\star}$, the crossing time of the source radius by the lens. However, if the source trajectory is tangential to the caustic line for a binary, the duration of such effects can be longer \citep{Hanetal2000}. 
\begin{table*}
	\centering
	\caption{The results from simulating single and binary microlensing events involving main sequence (MS) and giant (Giant) pulsating stars. The second and third columns, $\epsilon_{\rm{low}}$ and $\epsilon_{\rm{high}}$,are efficiencies for detecting pulsating signatures with low and high sensitivities, respectively. Here, $\left<f_{\rm b}\right>,~\left<A_{\rm{m}}\right>,~\left<\rm{SNR}\right>$ are the average blending parameter, maximum magnification factors as measured in $I$-band, and the signal-to-noise ratio (SNR) at the maximum magnification over the simulated events with detectable pulsation signatures, respectively.}
	\begin{tabular}{cccccccccccc}\toprule[1.2pt]
		& $\epsilon_{\rm{low}}[\%]$ &  $\epsilon_{\rm{high}}[\%]$ & $\left<\sigma_{\rm{m}}\right>/\left<\sigma_{\rm{base}}\right>$ & $\left<t_{\rm{s}}\right>$ & $\left< P\right>$ & $\left<\delta_{\rm{m}}\right>$ & $\left<f_{b}\right>$ & $\left<m_{\rm{base}}\right>$ & $\left<u_{0}\right>$ & $\left<A_{\rm{m}}\right>$& $\left<\rm{SNR}\right>$\\
		&  &  & & $\rm{(days)}$& $\rm{(days)}$ &  $\rm{(mmag)}$ & & $\rm{(mag)}$&  & & \\
		\toprule[1.2pt]\\
		\multicolumn{12}{ c }{$\rm{Single}~\rm{Microlensing}$}\\
		$\rm{Total}$& $50.89$ & $37.73$ & $0.84$ & $38.29$ & $7.03$ & $0.16$ & $0.73$ & $19.43$ & $0.46$ & $8.99$ & $85.6$\\
		$\rm{MS}$&$46.23$ & $33.83$ & $0.85$ & $39.20$ & $7.06$ & $0.16$ & $0.69$ & $19.92$ & $0.46$ & $8.22$ & $60.8$\\
		$\rm{Giant}$&$81.77$ & $63.55$ & $0.80$ & $34.88$ & $6.93$ & $0.14$ & $0.88$ & $17.63$ & $0.49$ & $11.85$ & $178.5$\\
		\hline\\
		\multicolumn{12}{ c }{$\rm{Binary}~\rm{Microlensing}$}\\
		$\rm{Total}$& $62.35$ & $49.30$ & $0.81$ & $39.26$ & $6.98$ & $0.14$ & $0.73$ & $19.58$ & $0.326$ & $129.65$ & $385.2$\\
		$\rm{MS}$& $57.96$ & $45.17$ & $0.82$ & $38.35$ & $6.94$ & $0.14$ & $0.69$ & $20.05$ & $0.322$ & $124.96$ & $255.8$\\
		$\rm{Giant}$& $92.21$ & $77.35$ & $0.76$ & $43.13$ & $7.17$ & $0.13$ & $0.89$ & $17.59$ & $0.344$ & $149.67$ & $937.9$\\
		\hline 
	\end{tabular}
	\label{tablast}
\end{table*}

Ignoring effects of transits and near-transits, the expectation is that the light curve, after dividing by the best-fitted simple microlensing model, will display the intrinsic periodic variability of the pulsating star. Of course, the pulsating modes exist in the baseline data as well; magnification by microlensing serves to decrease the photometric errors of the data points taken during lensing event, owing to the enhanced  brightness of the star. The improvement in photometric uncertainties by decreasing the stellar magnitude for real observations is shown in Figure \ref{errorpd}. These effects are apparent in the residuals panels of Figures \ref{fig2a} around $t$ of zero. The improved signal-to-noise enhances the probability of detecting stellar variability of low-amplitude pulsating stars, which is the topic of the next section.

\section{Observing stellar pulsations during microlensing}\label{varib}

To evaluate the effect of microlensing on detection of low-amplitude stellar pulsation, we calculate (i) the detection efficiency and (ii) the improvement in visibility of stellar pulsation, which are explained in the following.

\subsection{Detection~efficiency}  

Modeling of microlensing events proceeds generally as follows.  The first step is to assume simple microlensing models without consideration of second-order effects. Residuals are then analyzed, and any remaining trends are assessed in relation to established second-order effects. An example of a second-order effect is the parallax effect from Earth motion that can skew the simple microlensing model from being time symmetric about peak magnification \citep[see, e.g.,  ][]{1994Gould,parallax2019,parallaxII,parallax2020III, 2018Ryu}. 

However, pulsational variability is not one of the standard second-order effects that is considered, as variable stars have traditionally been excluded from potential stellar microlensing samples. Our goal here is to reassess the value of using variable stars, either to expand the pool of microlensing events, or as a tool for stellar astrophysics.

We adopt this same strategy in our theoretical study to determine the fraction of simulated events in which stellar pulsations will be detectable. To do so, we apply two main criteria. The first (hereafter, ``I.'') relates to the quality of the data, expressed as a threshold on the $\chi^{2}$ values of the residuals to indicate that second-order effects may be detected. The second (hereafter, ``II.'') is the practical consideration that the pulsation period cannot be longer than the characteristic timescale of the microlensing event.  Both criteria are quantified in what follows.

To test for the presence of stellar pulsations in lightcurves, we adopt for the first criterion (I.) $\Delta \chi^{2}_{\rm{lens}}>350$. Here, $\Delta \chi^{2}_{\rm{lens}}= \chi^{2}_{\rm{pulse}}-\chi^{2}_{\rm{nonpulse}}$, which is the difference in $\chi^{2}$ values from fitting the real model (microlensing of a pulsating source star, based on assumed parameters) and fitting a simple microlensing model for a non-pulsating star, $A_{\rm{best}}(t)$. The simple model for the non-pulsator uses the average observed flux of the pulsator from the baseline segments of the synthetic lightcurve, $\left<F_{\star, \rm{base}}\right>$. The fit from such simple models are plotted as the black dashed curves in Figures~\ref{lightcurve}, which is $A_{\rm{best}}(t)$.

\noindent This threshold amount (i.e., $350$) will include microlensing events with lower magnification factor (e.g., microlensing events that do not have obvious perturbations due to caustic-crossing features or transiting the source surface), but the stellar pulsations can be discerned (see, e.g., Figure \ref{fig2e}). However, this threshold is higher than those reported for planet detection in no caustic-crossing events \citep[see, e.g., ][]{Han2021dd}, because the time scales of planetary perturbations is shorter than that due to pulsation features of the source stars.

The above allows us to test for the presence of a second-order effect in the synthetic microlensing lightcurves. To ensure that periodic trends in the residuals can be detected, we additionally demand that the pulsating period be smaller than the magnification time scale, (II.) $P<t_{\rm{s}}$ In that case at least one pulsation cycle occurs during the lensing event. The magnification time scale, $t_{\rm{s}}$, is the time interval with the magnification factor larger than $3/\sqrt{5}=1.34$, which corresponds to the time interval in which the source star crosses the Einstein ring. Note that for single microlensing events, the timescale when the magnification factor exceeds $1.34$ is $t_{s}=2~t_{\rm E}\sqrt{1-u_{0}^{2}}$ (for events with $|u_{0}|<1$).  

\begin{figure*}
	\centering
	\includegraphics[width=0.49\textwidth]{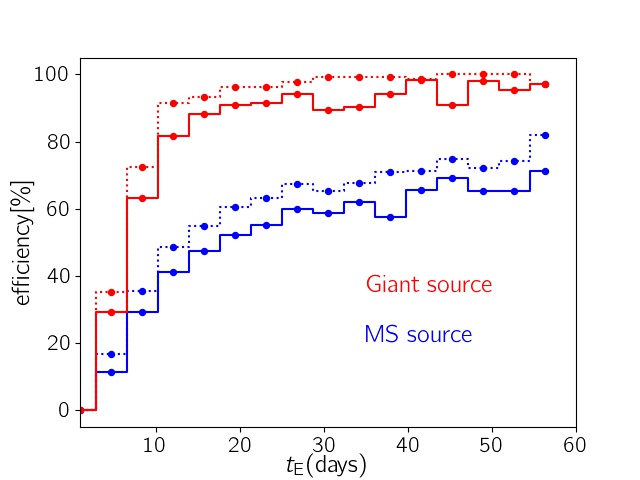}
	\includegraphics[width=0.49\textwidth]{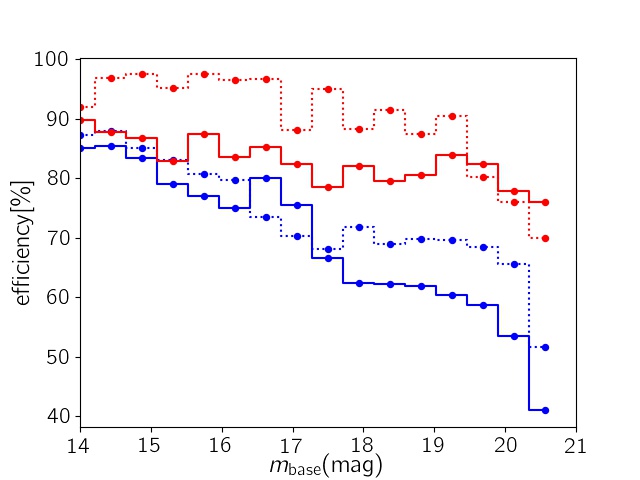}
	\includegraphics[width=0.49\textwidth]{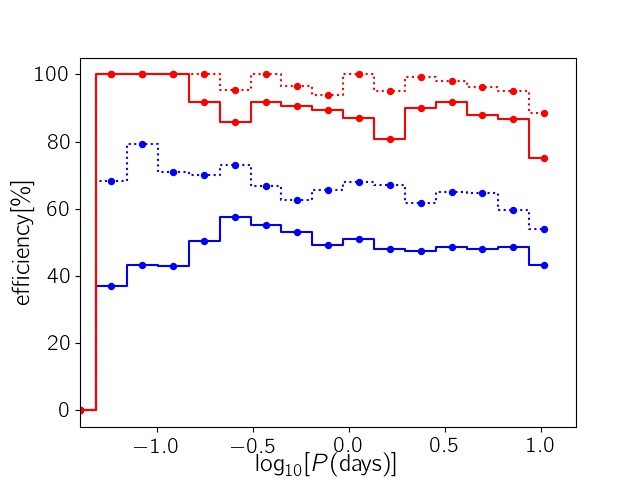}
	\includegraphics[width=0.49\textwidth]{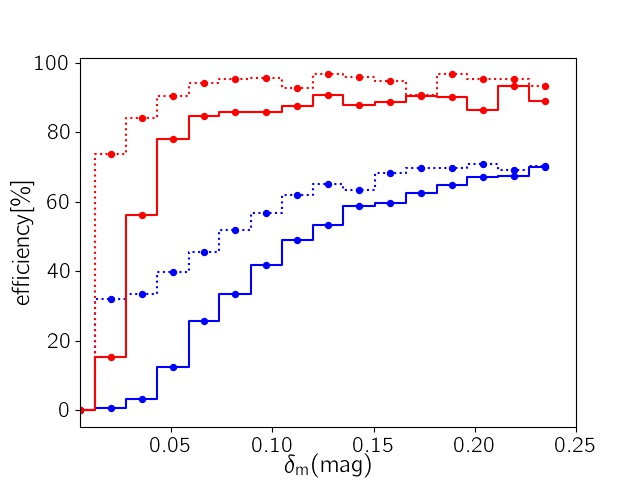}
	\caption{The efficiency for detecting stellar pulsation versus the Einstein crossing time, the apparent magnitude of the source star, the pulsating period and its amplitude. The solid and dotted curves are efficiencies for detecting pulsation signatures in single and binary lens events, respectively. The blue and red curves are for main-sequence (MS) and giants source stars.}
	\label{effiplot1}
\end{figure*}
We separately consider microlensing events involving main sequence (MS) and giant source stars. For each of them, by applying these conditions, we evaluate the efficiencies to discover stellar (radial and non-radial) pulsations in single and binary microlensing events. Here, detection efficiency refers to the fraction of the events with detectable pulsation signatures (i.e., referring to criteria I and II) to the total simulated events. These values are noted in the second column of Table \ref{tablast}. We do not separate the results for radial and non-radial pulsators, because they are similar for the same ranges of pulsation amplitudes and periods. We identify efficiencies for detecting stellar pulsations with low sensitivity, labelled $\epsilon_{\rm{low}}[\%]$ in the table. By considering two tighter criteria (higher sensitivity) as (I) $\Delta \chi^{2}_{\rm{lens}}>500$ and (II) $2~P<t_{\rm{s}}$ indicating a minimum of two pulsation cycles during the lensing, the detection efficiencies, $\epsilon_{\rm{high}}$, decrease, with values given in the third column of Table \ref{tablast}. The threshold of $500$ is usually considered as a robust criterion for planet detection in microlensing events \citep[see, e.g., ][]{Gould2010ww,Udalski2018ww}. 

\noindent Note that the table also provides the average parameter values for the pulsating source stars whose pulsation signatures are detectable with low sensitivity. Here, $\left<f_{\rm b}\right>$ is the average blending parameter over the simulated events with detectable pulsation signatures. The blending parameter $f_{\rm b}$ refers to the fraction of the source flux to the total fluxes entering into the source PSF. This parameter is unity if there is no blending stars close to the source star. $\left<A_{\rm{m}}\right>$ is the average of maximum magnification factors as measured in $I$-band over the simulated events with detectable pulsation signatures. The last column of this table is the average of the signal-to-noise ratio (SNR) at the maximum magnification over the events with detectable pulsating signatures. The SNR is evaluated as the square root of the number of photons entered to the source PSF \citep[see, e.g.,][]{2012Sajadian}:
\begin{eqnarray}
\rm{SNR}=\sqrt{t_{\rm{exp}} 10^{0.4 m_{\rm{zp}}} \left[ 10^{-0.4m_{\rm{base}}} A_{\rm{m}} + \Omega 10^{-0.4 \mu_{\rm{sky}}} \right] },
\end{eqnarray}
where, $t_{\rm{exp}}$ is the exposure time,  $\Omega$ is the angular area of the source PSF,  $\mu_{\rm{sky}}$ is the sky brightness and $m_{\rm{zp}}$ is the zero point magnitude. For the OGLE observation, $\Omega= 0.79~\rm{arcs}^{2}$, $t_{\rm{exp}}=120$ sec, $\mu_{\rm{sky}}= 19.7$ mag and $m_{\rm{zp}}=21.7$ mag  \citep{2015Wyrzykowski,sourcedis}. 


Generally, pulsations can be detected for events as long as $\left< t_{\rm s}\right> \sim 40$ days, which corresponds to $\left<t_{\rm E} \right> \sim 25.5$ days in single microlensing events (because in these events, we have $\left< t_{\rm s}\right> = \pi / 2~\left< t_{\rm E}\right>$).  In our simulations the average value of $t_{\rm E}$ over all simulated single events is $\sim 19.6$ days, which is shorter than that reported for OGLE microlensing events \citep[see, e.g.,][]{2019Mrozbulge,2015Wyrzykowski}. In fact, combining the OGLE and KMTNet data increases the probability of detecting short-duration events, and as a result decreases the averaged timescale of detectable events. The corresponding timescale for binary microlensing is more complicated, depending on the caustic configuration and the source trajectory.

Since giant source stars are on average brighter and the blending effect is reduced (see the last column of Table \ref{tablast}), the efficiency of detecting intrinsic pulsations is about twice that of main sequence stars.

For binary microlensing events, the simulated events mostly have caustic-crossing features. Additionally, in these events the time scale associated with the magnification threshold is somewhat longer than single-lens events.  For these reasons, their efficiencies are higher than those due to single events with similar amplitudes. However, it is the amplitude of the pulsations that dominates the detection efficiency. The pulsation features are mostly detectable in single and binary microlensing events with magnification peaks around $10$ and $130$, which result in SNR values of $\sim 80$ and $400$, respectively.

We plot in Figure \ref{effiplot1} the detection efficiency curves with the Einstein crossing time, the apparent magnitude of the source stars in the $I$-band, the period of the stellar pulsation and the pulsation amplitude. In each panel, four curves are given. Blue curves are efficiencies for detecting the pulsation signatures of main-sequence (MS) source stars, for single (solid) and binary (dotted) microlensing events. Red curves are for giant sources.

As expected, the probability for detecting pulsations is higher in longer microlensing events from intrinsically brighter source stars, e.g., caustic-crossing binary microlensing events from giant source stars. Higher pulsation amplitudes lead to larger detection efficiencies. We note that the efficiency for the amplitudes $0.05<\delta_{\rm m}<0.1$ mag is not zero. However, identifying these stellar pulsations with current surveys is almost impossible, but during the microlensing events the efficiencies for discerning these weak signatures are not zero and even reaches to $80\%$ in binary events from giant source stars.

The detection efficiency is maximized in the range of stellar periods $0.1\lesssim P \lesssim0.3$ days and it drops slowly with increasing period.  Variable stars with pulsation periods around this peak of detection efficiency include RR Lyrae stars, which are giants on the horizontal branch. These variable stars are radially pulsating and more than $10\times$ brighter than the Sun. Their pulsation amplitudes are generally larger than $0.03$ mag. $\delta$-Sct and $\gamma$-Dor pulsating stars are fainter than \rrl variable stars and have the pulsation periods less than 1~day  \citep{Poleski2010,Guzik2016}.  These intrinsic variable stars  can be discerned through microlensing with the highest efficiencies. 
\begin{figure*}
	\centering
	\subfigure[]{\includegraphics[angle=0,width=0.49\textwidth,clip=0]{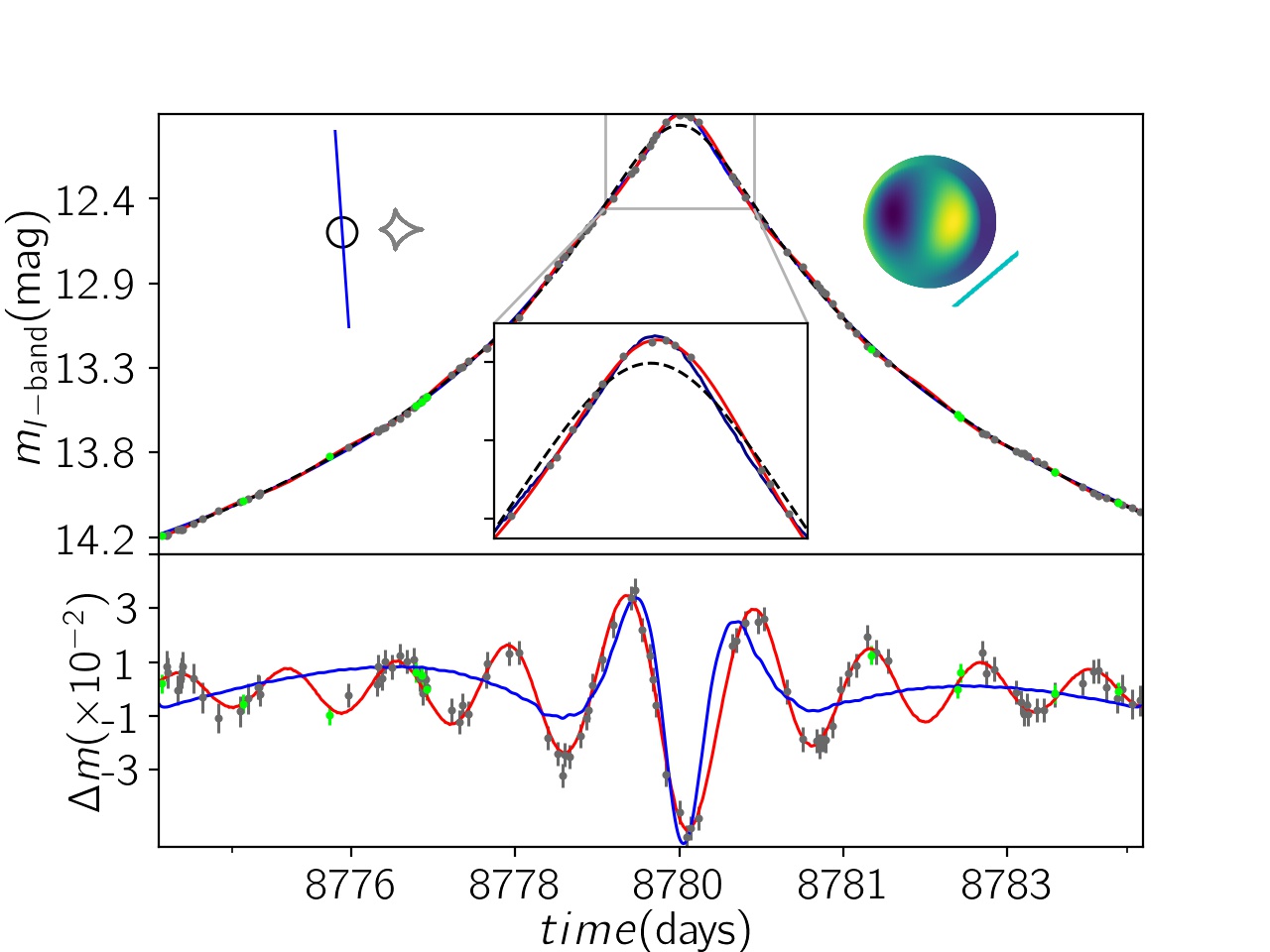}\label{light313}}
	\subfigure[]{\includegraphics[angle=0,width=0.49\textwidth,clip=0]{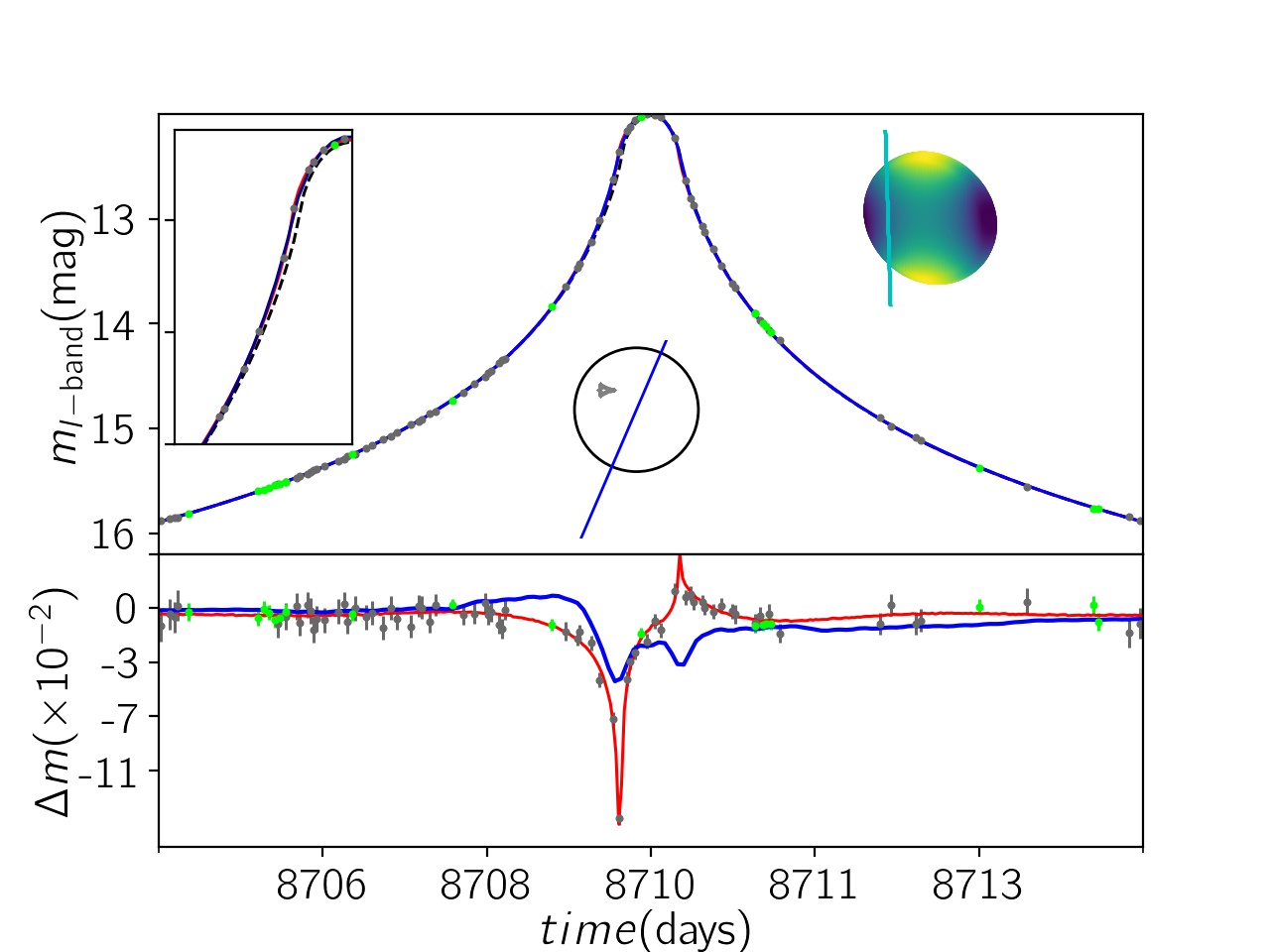}\label{light320}}
	\subfigure[]{\includegraphics[angle=0,width=0.49\textwidth,clip=0]{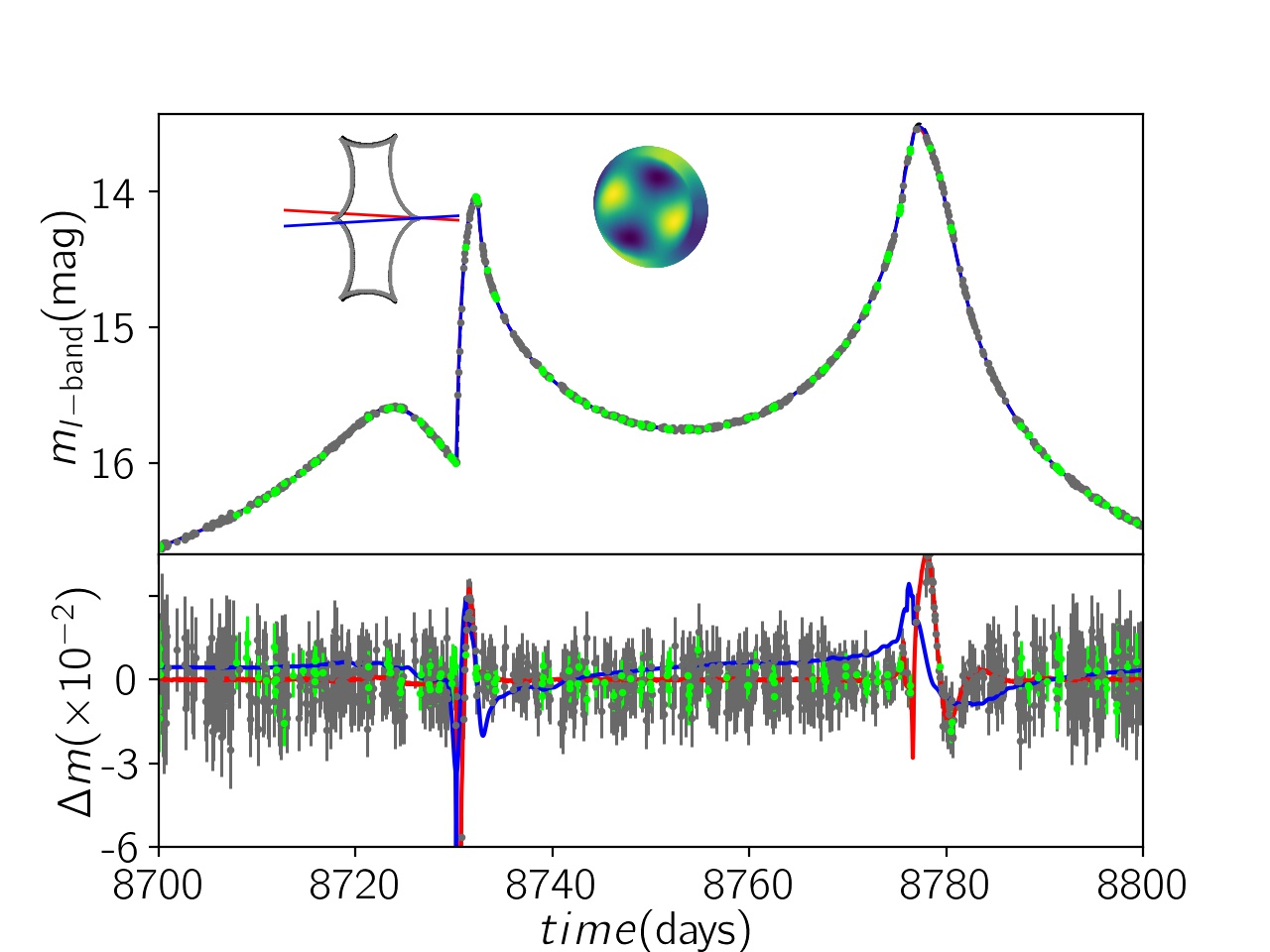}\label{light2004}}
	\subfigure[]{\includegraphics[angle=0,width=0.49\textwidth,clip=0]{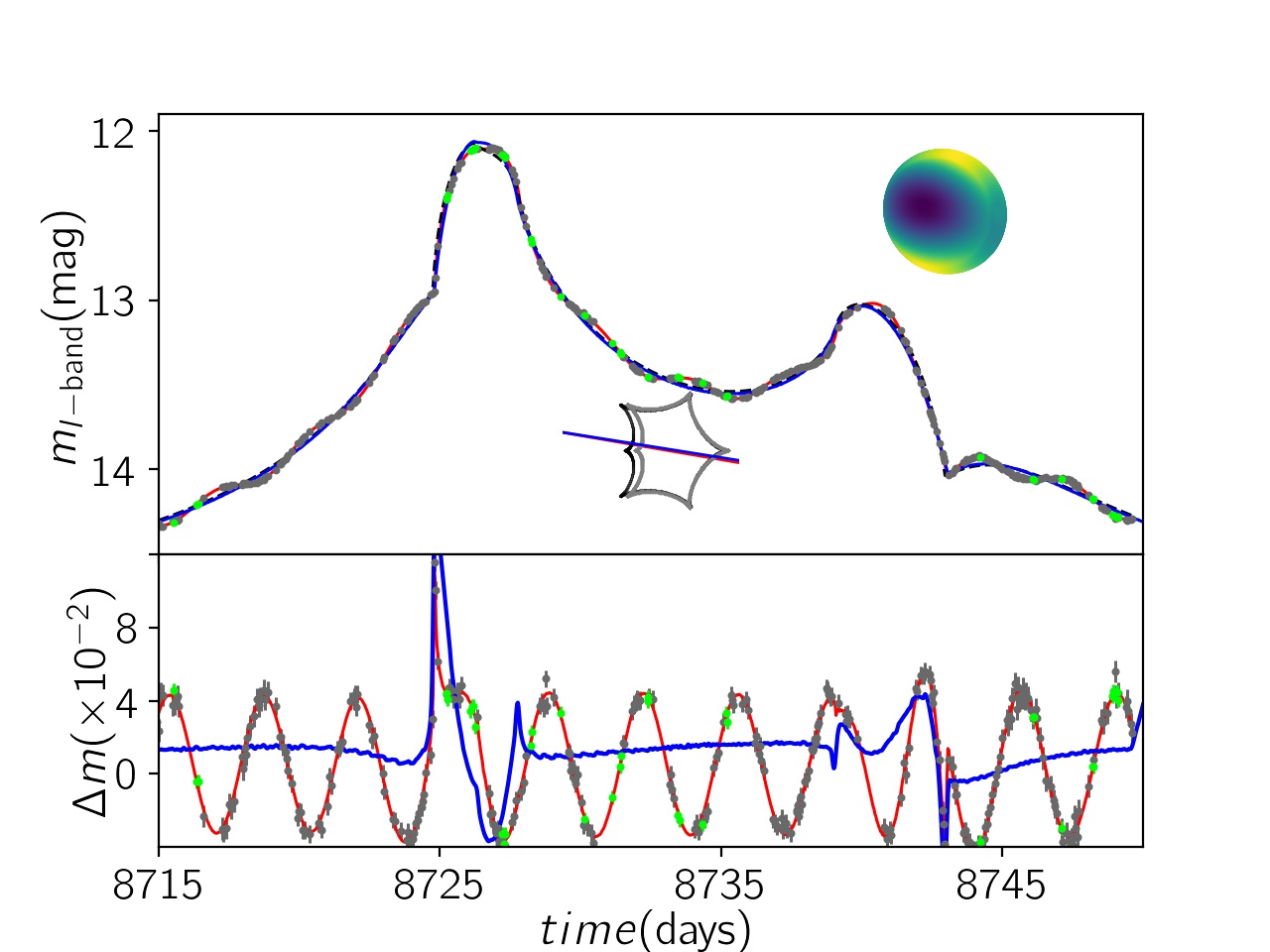}\label{light2}}
	\caption{Microlensing lightcurves from non-radially pulsating stars (red solid lines) and non-pulsating stars with average luminosity at the baseline (black dashed  curves). We show the best-fitted binary/planetary microlensing models for each light curve with solid blue curves. The insets are source trajectories (solid blue), caustic curves (gray curves) and the source disks (black circle) due to the best-fitted models. The (black) caustic curves and (red solid) source trajectories for the two bottom lightcurves are for the real models. Also, the gray caustic and blue source trajectory is for the best-fitting models. The chromatic schemes inside the lightcurves represent the variations of stellar temperature over their surfaces. In two top lightcurves, some parts of lightcurves with significant deviations with respect to black dashed curves are zoomed. The parameters for these lightcurves are given in Table \ref{tabmis}.}
	\label{missin}
\end{figure*}

Main-sequence stars can show periodic variability as well. The primary sources of periodic behavior can come from magnetic activity, stellar spots, eclipses, or planetary transits. The pulsation periods of these stars depend on stellar rotational speed or the separation between the binary components. Even though the shapes of these stellar variations are different from radial and non-radial pulsations (as simulated in this work), the detection efficiency through microlensing based on the definition in this work and $\Delta \chi^{2}$ values, should  depend on only their amplitudes
and periods. 

\subsection{Improvement in detectability of stellar pulsations}

Here, we study the effects of microlensing on the detection of stellar pulsations from another point-of-view. Decreasing the photometric errors during the lensing improves detectability of the stellar pulsations. To evaluate this improvement, we focus on the residuals of the light curve fits and the associated errors. We note that the photometric uncertainties in the synthetic data, $\sigma_{\rm m}$, decrease because of the light magnification during lensing. On the other hand, subtracting the best-fitted microlensing model from the data increases the uncertainty in each data point by the factor $\sqrt{2}$.



The improvement in detectability of stellar pulsations by lensing magnification depends on how much the photometric error bars improve during lensing. In this regard, we calculate the ratio of the averaged photometric error bar during lensing to that from the baseline amount, $\left< \sigma_{\rm m} \right>/\left< \sigma_{\rm{base}} \right>$. The smaller this ratio, the higher the reliability for discerning the stellar pulsations. The average value of this ratio over all simulated events is $\sim 0.75-0.85$. Values for stars with (radial and non-radial) pulsations with single or binary microlensing events are given in the fourth column of Table \ref{tablast}. Decreasing the photometric error bars is the main reason for higher efficiency in detecting stellar pulsations during lensing events, because SNR is improved during a lensing event. The relation between the photometric error bars and the SNR is 
\begin{eqnarray}
\sigma_{\rm m}= -2.5 \log_{10}\left[1+ \frac{\delta F(t)}{F(t)}\right] = - 2.5 \log_{10}\left[1+\frac{1}{\rm{SNR}}\right], \label{snr}
\end{eqnarray}
\noindent where $F(t)$ and $\delta F(t)$ are total flux and its fluctuation at any given time. Assuming $\left<\sigma_{\rm m}\right> =0.8\left<\sigma_{\rm{base}}\right>$ and using Equation \ref{snr}, 
the SNR improves $\simeq 25\%$ (i.e., $\rm{SNR} \simeq 1.25~\rm{SNR}_{\rm{base}}$).

With real observational data, the true interpretation of perturbations in the residuals depends on many factors. For instance, if the period of pulsation is comparable to the magnification time scale, the periodic features of the perturbations may be missed. Also, in high-magnification microlensing events involving a transit by a
single lens, or during caustic-crossings, the magnification curve does not follow the standard simple shape with a non-variable source, which likely causes wrong best-fitted models. In the next section, we discuss the prospect of misinterpreting the microlensing of pulsating source stars in terms of other second-order effects.

\section{Misinterpreting pulsation-induced perturbations}\label{miss}

Some non-radially pulsating source stars have very small pulsation amplitudes, although the amplitudes of variation in their temperature and radius are large. But overall their stellar luminosities may not change significantly. For instance, when the stellar polar axis is toward the observer and $m>0$, variations in the stellar surface
temperatures are averaged across the face of the stellar surface. The result is very small variation of stellar luminosity as seen by the observer \citep[see, ][]{PaperII}. But if in this orientation, a single lens were to transit across the face of surface or if the source star passes a caustic curve, there could be detectable perturbations to the residuals from a standard microlensing fit, but they would not be periodic and might be attributed to some other effects. If detected, the perturbations might be misattributed to some other second-order effect.

Figure \ref{missin} displays some examples of the events just described. In each panel, the real lightcurves in the $I$-band magnitude are shown with red solid curves. The real lightcurves from non-pulsating source stars are shown with black dashed curves. For each lightcurve, a best-fit binary or planetary microlensing model is displayed (blue solid curve) that assumes non-pulsating source stars and a search within a grid of predetermined models, as developed by C.~Han \citep[see, e.g., ][]{Han2020a,Han2020b} \footnote{\url{http://astroph.chungbuk.ac.kr/~cheongho/}}. These fitted models are shown with blue solid curves. In the residuals panels, we show the deviation of red and blue solid lightcurves with respect to the black dashed ones (real model without pulsation features).

The input parameters for the real microlensing lightcurves are given in Table \ref{tabmis}. For each lightcurve, the parameters of the best-fitted models are given in the second row. In this table, $\xi$ is the angle between the binary axis and source trajectory in binary lensing events, and the last column ``$\rm{No.}$'' is the number of their synthetic (OGLE and KMTNet) data points. The number of data points can be used to evaluate the normalized $\chi^{2}$ values. 

The first two lightcurves are single lens and high-magnification microlensing events from non-radially pulsating source stars with amplitudes of less than $1$ mmag. In lightcurve \ref{light313}, the periodic features around the peak of the light curve model arise from the close passage by a small central caustic due to a wide binary system. However, the best-fitting model compensates only for two pulsation features around the peak, its $\chi^{2}$ value is larger than the real model by only $\sim 800$. We note that the number of data points for this lightcurve is $105$.

The second lightcurve, \ref{light320}, is a single lens transit microlensing event also involving a non-radially pulsating source star with small pulsation amplitude. While the lens is transiting the source surface (i.e., $u < \rho_{\star}$ where $u$ is the lens-source distance normalized to the Einstein radius), the lightcurve is not symmetric in time owing to the pulsational modulation of the temperature distribution across the surface. The asymmetric feature in the lightcurve can be seen by zooming around the time of peak magnification.  In the modeling process, this asymmetric feature results from crossing over a caustic curve that is more compact than the source size (shown as the left insets), which can arise from a low-mass companion around the microlens. However, the difference between $\chi^{2}$ values is high, because the data points around the peak are not well reproduced with the best-fitted model. The number of data points in this lightcurve is $369$.
\begin{table*}
	\centering
	\caption{The real parameters of microlensing lightcurves from non-radial pulsating stars shown in Figure \ref{missin}. For each lightcurve, the parameters of the best-fitted models are given in the second row. $\xi$ is the angle between the binary axis and source trajectory in binary lensing events. The last column $\rm{No.}$ is the number of their synthetic (OGLE and KMTNet) data points.} 
	\begin{tabular}{cccccccccccccc}\toprule[1.2pt]
		$\rm{Fig.~No.}$& $\delta_{\rm}$ & $P$ &  $i$  & $l$ & $m$ & $u_{0}$ & $\rho_{\star}$ & $t_{\rm{E}}$  & $\chi^{2}$ & $q$ & $d$ & $\xi$ & $\rm{No.}$\\
		& $\rm{(mmag)}$  & $\rm{(days)}$ & $\rm{(deg)}$ & & & & &$\rm{(days)}$  & & & $\rm{(R_{E})}$& $\rm{(deg)}$ & \\	
		\toprule[1.2pt]
		\ref{light313}& $0.46$ & $1.21$ & $89.3$ & $3$ & $1$ & $0.013$ & $0.009$ & $65.3$  & $131.1$ & $--$ & $--$ & $--$ & $105$\\
		$\rm{Best}$-$\rm{fitted}$& $--$ & $--$ & $--$ & $--$ & $--$ &  $0.027$ & $0.008$ &  $26.9$ & $928.6$ & $0.23$ & $5.86$ & $86.0$ & $105$ \\
		\hline
		\ref{light320}& $0.53$ & $2.82$ & $89.5$ & $2$ & $2$ & $0.004$ & $0.006$ & $72.9$  & $ 2738.5$ & $--$ & $--$ & $--$ & $369$ \\
		$\rm{Best}$-$\rm{fitted}$& $--$ & $--$ & $--$ & $--$ & $--$ &  $ 0.004$ & $ 0.006$ &  $70.6$ &  $8185.1$ & $0.001$ & $2.3$ & $293.3$ & $369$ \\
		\hline
		\ref{light2004}&$0.48$ & $6.43$ & $89.6$ & $4$ & $2$ & $0.023$ & $0.011$ & $77.8$ &  $4086.9$ & $0.77$ & $0.99$ &  $-3.3$ & $1269$\\
		$\rm{Best}$-$\rm{fitted}$& $--$ & $--$ & $--$ & $--$ & $--$ &  $ 0.023$ & $ 0.012$ &  $76.3$ &   $6085.7$ & $0.76$ & $1.01$ & $3.4$ & $1269$\\
		\hline
		\ref{light2}& $44.10$ & $3.37$ & $40.7$ & $2$ & $1$ & $0.021$ & $0.064$ & $22.7$ & $1294.3$ & $0.29$ & $1.17$ &  $-9.72$ & $398$\\
		$\rm{Best}$-$\rm{fitted}$& $--$ & $--$ & $--$ & $--$ & $--$ & $0.031$ & $0.057$ &  $24.4$ &   $25955.0$ & $0.25$ & $ 1.12$ & $-9.06$ & $398$ \\
		\hline
	\end{tabular} \label{tabmis}
\end{table*}

The other two lightcurves are for binary microlensing events of non-radially pulsating source stars. In the plots, the inset represent the real models of caustic curve (black one) and the source trajectory (solid red line). Additionally, the best-fitted models of caustic curve (gray one) and the source trajectory (blue solid line) are plotted over them. The best-fitting models are very close to the real ones. In these events, the stellar pulsations produce large deviations at the times of caustic crossings. In the right-hand lightcurve, \ref{light2}, these deviations around the first peak tend to alter the inferred mass ratio, $q$.

\section{conclusions} \label{conclu}

Several microlensing events of variable source stars have been reported. In all of these events, the source stars were obviously variable at the baseline. In such events, measuring the stellar variability leads to indications of the source type, its distance and radius. By knowing the source properties, one can determine the lens parameters including its mass and distance from the observer. A significant benefit from stellar pulsations is the prospect of resolving the microlensing degeneracy and determining the  physical parameters of the lenses.

However, it is possible that the pulsational amplitudes of the variable source stars are relatively small, such that the pulsational behavior is not apparent in the baseline data (i.e., before and after a microlensing event).  Yet the pulsations could affect the microlensing lightcurves. For these events, there are two main points. First is whether the lensing effect could help discern the presence of pulsational variability.  Second is whether the perturbations in the microlensing lightcurves arising from the low-amplitude stellar pulsations are distinguishable from other effects that might perturb the lightcurves.  This paper has reported on studies of both of these issues.

Microlensing decreases photometric uncertainties of the data points by increasing the signal-to-noise ratio owing to the enhanced brightness of the lensed source. This improvement raises efficiencies for detecting the pulsating signatures from variable stars. We simulated OGLE and KMTNet observations of ongoing microlensing events from low-amplitude ($\delta_{\rm m} \in [0.005,~0.25]$ mag) radially and non-radially pulsating source stars. In this regard, we determined the cadences and photometric error bars for synthetic data points based on real observations by these surveys. This Monte-Carlo simulation of single and binary microlensing events revealed that on average, uncertainties during the microlensing event decrease to around $0.75-0.85$ of their amounts at the baseline. Also, efficiencies for detecting the stellar pulsations from low-amplitude pulsations are around $50-60\%$ with the criterion that $\Delta \chi^{2}>350$ and a pulsation period shorter than the magnification time scale. By considering criteria with higher sensitivity ($\Delta \chi^{2}>500$ and at least two pulsation cycles over the magnification time scale), the efficiency is around $40-50\%$. Efficiencies for the giant stars are as twice as large as for the main-sequence stars.

Although microlensing can increase the detectability of the stellar pulsations, some lightcurves of these pulsating source stars do not  have obvious periodic features in their residuals (after subtracting the best-fitted simple microlensing model).  In these cases the perturbations arising around peak magnification could be misinterpreted
with alternate sources of perturbations in microlensing events. We have shown example lightcurves in section \ref{miss} to find that small asymmetric perturbations around microlensing peak could be misidentified in terms of the source star crossing over or passing close to a caustic curve, in which the caustic is smaller than the source disk. Not taking pulsations into account could misattribute such a signature as arising from planetary systems or wide/close binary ones.

\section*{Acknowledgements}
S. Sajadian thanks the Department of Physics, Chungbuk National University. She specially thanks C.~Han and D.~Kim to teach how to find the best-fitted binary and planetary microlensing events. R. Ignace acknowledges support for this work from the National Science Foundation under Grant No. (AST-1747658).  We thank the referee for the helpful comments and good suggestions. 

\section*{DATA AVAILABILITY}
The data underlying this article will be shared on reasonable request to the corresponding author.

\bibliographystyle{mnras}
\bibliography{references}
\end{document}